\documentclass[sigconf,screen]{acmart}
\AtBeginDocument{%
  \providecommand\BibTeX{{%
    \normalfont B\kern-0.5em{\scshape i\kern-0.25em b}\kern-0.8em\TeX}}}

\setcopyright{acmlicensed}
\acmPrice{15.00}
\acmDOI{10.1145/3549035.3561185}
\acmYear{2022}
\copyrightyear{2022}
\acmSubmissionID{fsews22msrpsmain-p94-p}
\acmISBN{978-1-4503-9457-4/22/11}
\acmConference[MSR4P\&S '22]{Proceedings of the 1st International Workshop on Mining Software Repositories Applications for Privacy and Security}{November 18, 2022}{Singapore, Singapore}
\acmBooktitle{Proceedings of the 1st International Workshop on Mining Software Repositories Applications for Privacy and Security (MSR4P\&S '22), November 18, 2022, Singapore, Singapore}

\usepackage{natbib}
\usepackage{multicol,multirow}
\usepackage{csquotes}
\usepackage[nointegrals]{wasysym} 
\usepackage{kantlipsum}
\usepackage[english]{babel}
\usepackage{graphicx}
\usepackage{hyperref}
\usepackage[linesnumbered,ruled]{algorithm2e}

\makeatletter
\providecommand*{\Dashv}{%
  \mathrel{%
    \mathpalette\@Dashv\vDash
  }%
}
\newcommand*{\@Dashv}[2]{%
  \reflectbox{$\m@th#1#2$}%
}
\makeatother

\usepackage{xcolor}

\usepackage[shortlabels]{enumitem}

\newcommand{\Answer}[1]{%
\item[\bfseries A\arabic{enumi}] #1
}

\begin{document}

\title{Assessing Software Privacy using the Privacy Flow-Graph}

\author{Feiyang Tang}
\affiliation{%
  \institution{Norwegian Computing Center}
  \city{Oslo}
  \country{Norway}}
\email{feiyang@nr.no}

\author{Bjarte M. Østvold}
\affiliation{%
  \institution{Norwegian Computing Center}
  \city{Oslo}
  \country{Norway}}
\email{bjarte@nr.no}

\renewcommand{\shortauthors}{Tang and Østvold}

\begin{abstract}

    We increasingly rely on digital services and the conveniences they provide. Processing of personal data is integral to such services and thus privacy and data protection are a growing concern, and governments have responded with regulations such as the EU's GDPR.
    Following this, organisations that make software have legal obligations to document the privacy and data protection of their software. This work must involve both software developers that understand the code and the organisation's data protection officer or legal department that understands privacy and the requirements of a Data Protection and Impact Assessment (DPIA).

    To help developers and non-technical people such as lawyers document the privacy and data protection behaviour of software, we have developed an automatic software analysis technique. This technique is based on static program analysis to characterise the flow of privacy-related data. The results of the analysis can be presented as a graph of privacy flows and operations---that is understandable also for non-technical people.  
    We argue that our technique facilitates collaboration between technical and non-technical people in documenting the privacy behaviour of the software. We explain how to use the results produced by our technique to answer a series of privacy-relevant questions needed for a DPIA.
    To illustrate our work, we show both detailed and abstract analysis results from applying our analysis technique to the secure messaging service Signal and to the client of the cloud service NextCloud and show how their privacy flow-graphs inform the writing of a DPIA.

\end{abstract}



\maketitle

\section{Introduction}
Privacy has been widely discussed in recent years --- with the rise in public awareness and associated legislative developments, guaranteeing privacy while processing large amounts of private user data has become an important topic. Following recent law implementations such as the GDPR, we now have a regulated and clear framework for ensuring privacy compliance, which mandates documenting software properties through, for example, a Data Privacy Impact Assessment (DPIA). Such an examination must include all parts of the software and it requires a grasp of the software as well as sufficient technical knowledge to analyse the implementation. As a result, we would anticipate a development team expert who has a brief grasp of the implementation while also having sophisticated analysis and tools at their disposal to assist ensure that critical questions in evaluation frameworks such as DPIA can be answered. 

The reality, however, is considerably different. While having a privacy compliance checking process operating alongside a software development life cycle is important, analysis and tools at the code level with tailored assistance to legal experts are insufficient. In the meantime, DPIA questions require an understanding of both technical and legal aspects. This means that performing a successful DPIA cannot be done exclusively by a non-technical Data Protection Officer (DPO) who specialises in data protection policy or a technical professional from the data controller (e.g., a developer in the service provider organisation) with programming experience. Simultaneously, it is difficult for developers to keep track of every single change in terms of private data processing among hundreds of lines of code.

This raises the following question: how can we help both technical developers (from or work for data controllers) and non-technical (DPOs) individuals examine privacy compliance in software? Since tracking the flow of data originating from users is important for privacy protection, we must check sensitive user inputs to the software and use an explainable abstraction to illustrate the privacy behaviours in the software, address privacy elements, and provide assistance in producing a better privacy analysis.

We propose privacy flow-graphs as a means to help both developers and DPOs, they can adopt our technique to discover privacy-related behaviours in software.
Such graphs produced by our technique enable documenting private data processing actions, assist organisations (the data controller) in showing compliance with their duties and assist the DPO in carrying out its missions. Illustrating the processes may also assist developers to construct more privacy-compliant software and achieve privacy-by-design throughout development and deployment.

Our contributions are:
\begin{itemize}[noitemsep,topsep=0pt]
    \item The definition of the privacy flow-graph (Section~\ref{sec:privacy-flow-graph})
    \item How to write a DPIA informed by the privacy flow-graph (Section~\ref{sec:assessing}).
    \item A static program analysis that builds the privacy flow-graphs for Java programs (Section~\ref{sec:analysis}).
\end{itemize}
We demonstrate the utility of our research by examining privacy-related trends in two well-known Java applications: Signal and NextCloud (Section~\ref{sec:experiment}).

\section{Motivation}
Examining data protection compliance is essential for the vast majority of software released to the market, as well as for every service update when new user data must be analysed or when the way data is handled changes. Legal regulations such as the GDPR necessitate that legal experts obtain detailed privacy-related information processes from software developers. This implementation-specific information is typically obtained through manual labour by developers, and may not include all that a legal expert needs.

However, there are developers that are unfamiliar with the existing software and might have difficulties providing in-depth information to legal experts.

This circumstance motivated us to design a lightweight, semi-automated program analysis technique that automatically analyses how and where personal data is accessed and processed, therefore providing software developers and DPOs with a great deal of ease.

\section{Preliminaries}
In this section, we describe the preliminary aspects of our analysis: the local and global data-flow, the privacy flow-graph, the source and sink methods, and the handcrafted datasets we created to support the analysis.

Let $c,d$ denote classes, $n,m$ methods, and let notation $c.m$ make explicit that class $c$ that declares $m$.
We assume that method names are unique in a class.

\subsection{Local data-flow in methods}
\newcommand{\methodstart}{\ensuremath{\mathtt{start}}}
\newcommand{\returnstmt}{\ensuremath{\mathtt{return}}}
\newcommand{\invokestmt}[1]{\ensuremath{\mathtt{invoke}\ #1}}
\newcommand{\inputprim}{\ensuremath{\mathtt{i\_primitive}}}
\newcommand{\outputprim}{\ensuremath{\mathtt{o\_primitive}}}
We define some notation to refer to results obtainable from the control flow graph (CFG) of a method.
These results concern the kind of values that may flow between various points either inside the method body.
\begin{definition}[Method data-flow point $p$]
  A data-flow point $p$ associated with a method $c.m$ is one of the following:
  \begin{quote}
    {\methodstart} -- the start of the method;\\
    \invokestmt {d.n}\,$_i$ -- an invocation of method $d.n$;\\
    \inputprim$_i$ -- an input primitive;\\
    \outputprim$_i$ -- an output primitive;\\
    \returnstmt$_i$ -- a return statement.
  \end{quote}
\end{definition}
\newcommand{\localflow}[2]{\ensuremath{#1 \mapsto #2}}
\newcommand{\cfg}[1]{\ensuremath{\mathit{CFG}(#1)}}
\newcommand{\entailed}[1]{\ensuremath{\Dashv \cfg {#1}}}
\newcommand{\localflowEntailed}[3]{\ensuremath{(\localflow{#1}{#2}) \entailed{#3}}}
\newcommand{\bFlow}[1]{\ensuremath{\mathit{begin}(#1)}}
\newcommand{\eFlow}[1]{\ensuremath{\mathit{end}(#1)}}
\begin{definition}[Local data-flow $F$; beginning, end]
  Let $p, p'$ be data-flow points, let $F$ be $\localflow p {p'}$ and let $c.m$ be a method.
  We write $F \entailed {c.m}$ to means that the control-flow-graph of $c.m$ specifies a \emph{local data-flow} $F$, that is, that values may flow from $p$ to $p'$.
  We refer to $p$ as the \emph{beginning} of $F$, denoted $\bFlow F$ and $p'$ as the \emph{end} of $F$, denoted $\eFlow F$.
\end{definition}
An invocation can be both a beginning and an end of a flow, whereas the start of the method and an input primitive can only be a beginning, and a return statement and an output primitive can only be an end.

We are concerned with all data-flows that originate from the use of an input primitive.
We now define some particular types of flows.
\newcommand{\oflow}{\ensuremath{F^o}}
\begin{definition}[Source flow, $\oflow$]
  Given method $c.m$ where \\ \localflowEntailed {\inputprim_i} {\returnstmt_j} {c.m}.
  This flow  is called a \emph{source flow}, denoted $\oflow$.
\end{definition}
\newcommand{\iflow}{\ensuremath{F^i}}
\begin{definition}[Sink flow, $\iflow$]
  Given method $c.m$ where \\ \localflowEntailed \methodstart {\outputprim_i} {c.m}.
  The flow is called a \emph{sink flow}, denoted $\iflow$.
\end{definition}

\subsection{Global data-flow \& the privacy flow-graph}
\label{sec:privacy-flow-graph}
We now consider global data-flow, specifically data-flows between methods of different classes, those are, all data-flows that start from the use of an input primitive.

We extend the concept of a data-flow from local flows $F$ inside methods to global flows $G$ across methods.
A global data-flow is defined by a series of local data-flows, each corresponding to a method invocation, and that satisfies certain conditions.
\begin{definition}[Global data-flow $G$]
  A \emph{global data-flow} $G$ is finite series of two or more local data flows,
  $F_1 \cdots F_n$.
  The notions of beginning and end extend to $G$ in an obvious way.
  Furthermore, any $F_k, F_{k+1}$ above must satisfy the following:
  Let $c_k.m_k$ be such that  $F_k \entailed {c_k.m_k}$ and $c_{k+1}.m_{k+1}$ such that $F_{k+1} \entailed {c_{k+1}.m_{k+1}}$ and $\eFlow {F_k} = \returnstmt_i$ and $\bFlow {F_{k+1}} = \invokestmt {{c_k.m_k}}_j$ for some $i,j$.
\end{definition}

A global data-flow $G=F_1 \dots F_n$ is a \emph{privacy flow} if $F_1$ is a source flow.
We are especially interested in global data-flows that involve data from input primitives ending up in output primitives.

Let $P$ be a program with privacy flows $G_1, \dots, G_n$. The \emph{privacy flow-graph} is a graph where there the nodes are all methods involved in a privacy flow and the edges are pairs of methods involved in successive flows $F_k,F_{k+1}$ part of some $G_j$.

\subsubsection{Java specifics}\label{sec:javaspec}
Here we consider some issues in adapting our data-flow definitions to Java.

First, we define rich types with the intuition that we are only interested in flows that involve values of these kinds of types.

\begin{definition}[Rich type]
A \emph{rich type} is any of following: the primitive data types \texttt{string}, \texttt{int}, \texttt{byte}, the object types, as well as arrays of rich types.
\end{definition}
Values of rich types are those values that may contain privacy-related information. In principle, a \emph{boolean} could also be relevant to privacy, but we limit our scope to the rich types to simplify our task. We are concerned with the processing of privacy-related data and not with the leakage of bits of privacy information stemming from such processing. 

All non-trivial programs refer to either standard libraries or third-party libraries and thus source flows and sink flows may take place inside the methods of these libraries.
In order to include these flows without analyzing the libraries, we introduce the concept of source methods and sink methods where such flows happen, and we apply a separate library analysis to pre-build a collection of source and sink methods.

\newcommand{\om}{\ensuremath{\mathit{om}}}
\newcommand{\im}{\ensuremath{\mathit{im}}}
A \emph{source method} is a method whose invocation results in a source flow, and we denote it as $\om$.
A \emph{sink method} is a method whose invocation results in a sink flow, denoted $\im$.

\subsubsection{Library analysis}
We have manually constructed a dataset of source and sink methods in the native Java library\footnote{Based on JDK 8u201} as well as the most used third-party Java libraries across different categories\footnote{Jackson, Log4j2, Apache Commons, Guava, HttpClient, JMS, Joda Time, Apache MINA, Apache Commons Codec and Derby}. The third-party libraries were selected from the Maven Repository list based on their download frequency\footnote{Maven Repository: \url{https://mvnrepository.com/}}. There are 158 Java source methods and 257 third-party library methods, which are divided into five groups based on the return data type. Table~\ref{tab:om} displays three Java source method samples and three from third-party libraries.
\begin{table*}[ht]
\centering
\caption{Examples of source methods\label{tab:om}}
\begin{tabular}{lll}
\textbf{Method signature}                                 & \textbf{Category}  \\ \hline\hline
int java.io.DataInputStream.read(byte[])     & I/O      \\
java.lang.String java.net.URL.getQuery()             & Network \\
java.sql.ResultSet java.sql.Statement.getResultSet() & Database \\
int org.apache.commons.io.input.ProxyInputStream.read(byte[])     & I/O      \\
org.apache.http.ssl.SSLContextBuilder org.apache.http.ssl.SSLContextBuilder.loadKeyMaterial()  & Network \\
java.sql.ResultSet org.apache.derby.iapi.jdbc.BrokeredStatement.executeQuery(java.lang.String) & Database
\end{tabular}%
\end{table*}

Similarly, we created a dataset that included 350 sink methods from the same Java and 365 sink methods from the third-party libraries we investigated for the source method. Five examples of sink methods are displayed below in Table~\ref{tab:im}.

\begin{table*}[ht]
\centering
\caption{Examples of sink methods\label{tab:im}}
\begin{tabular}{lll}
\textbf{Method signature}                                 & \textbf{Category}  \\ \hline\hline
void java.util.logging.Logger.log(java.util.logging.LogRecord) & Log \\
void java.io.BufferedWriter.write(int)     & I/O      \\
void javax.servlet.http.HttpServletResponse.sendRedirect(java.lang.String)      & Network \\
void com.sun.xml.txw2.output.XMLWriter.comment(char[],int,int) & I/O \\
java.net.HttpURLConnection org.jsoup.helper.HttpConnection(org.jsoup.Connection) & Network       
\end{tabular}%
\end{table*}

A global privacy data-flow is made up of many nodes that represent various methods. Different methods imply different types of data processing; to help demonstrate these processes, we characterise \emph{process} under four categories.

\begin{definition}[Process]
A process is a local data-flow $F$ in a \emph{privacy flow} $G=F_1 \dots F_n$ that is not a source flow $F^o$ or a sink flow $F^i$. \end{definition}

To specify some special kinds of processes, we use the following separate terms:
\begin{itemize}
    \item Security process, if a process involves cryptography, database, security, or network packages.
    \item Authentication process, if authentication is involved.
    \item Initialisation process, if a process initialises a class.
    \item Non-privacy process, if it does not belong to either of the three categories above.
\end{itemize}

\section{Assessing data privacy}
\label{sec:assessing}
It is challenging for software developers and legal privacy experts to have a mutual understanding and benefit from each other's expertise and insights.
To address this, we examine how to leverage information from data flows in software to answer particular concerns related to GDPR rules.
According to Article 4 in GDPR, \textit{``the data controller determines the purposes for which and the means by which personal data is processed''}; hence, software providers (organisations) are data controllers if the organisation develops its own software. Otherwise, the software developers provide the implementation to the data controllers who are responsible for privacy protection. In this paragraph, we first look at the core GDPR obligations of the data controller, which serves as the duty of DPOs, and then discuss how we may help DPOs answer key DPIA questions (the document created by the approach in this study is referred to as a DPIA.).

\subsection{Obligation of the data controller}
Article 24 in the GDPR~\cite{european_commission_regulation_2016} states several obligations of the data controller which should be monitored by the DPO:
\begin{itemize}
    \item by default and by design, the data controller should have a record of processing activities (Article 30);
    \item to ensure the security of the processing (Article 32);
    \item to notify personal data breaches to the supervisory authorities (Article 33);
    \item to communicate personal breaches to the data subject (article 34)
    \item to conduct DPIA (Article 35);
    \item to conduct prior consultation with supervisory authorities (Article 36).
\end{itemize}
The DPOs' role is to monitor whether the data controller fulfilled all of their commitments, which includes performing a DPIA when required. The writing of a DPIA is a shared duty for data controllers and DPOs.
 
As one of the major data protection authorities in Europe, the Irish Data Protection Commission~\cite{DataProt26:online} provides a short explanation of what DPIA contains:
\begin{displayquote}
``\textit{A DPIA describes a process designed to identify risks arising out of the processing of personal data and to minimise these risks as far and as early as possible.}''
\end{displayquote}
Here we picked one of the most often used sample templates for generating a DPIA from 
the British Information Commissioner's Office (ICO)~\cite{ICO}.

Under \textit{Section 2: Describe the processing} of the template, there are three questions:
\begin{itemize}
    \item Describe the nature of the processing: how will you collect, use, store and delete data? What is the source of the data? Will you be sharing data with anyone? You might find it useful to refer to a flow-graph or another way of describing data flows. What types of processing identified as likely high risk are involved?
    \item Describe the scope of the processing: what is the nature of the data, and does it include special category or criminal offence data? How much data will you be collecting and using? How often? How long will you keep it? and more
    \item Describe the context of the processing: what is the nature of your relationship with the individuals? How much control will they have?
\end{itemize}
Also under \textit{Step 5: Identify and assess risks}, DPIA requires \textit{``Describe the source of risk and nature of the potential impact on individuals.''}

With a list of privacy data-flows listed under different categories, developers and DPOs could identify the parts of the program that collect privacy data from users and the relevant risky sinks. As a result of identifying privacy flows, they can pinpoint exposure risks and offer solutions to minimise those risks.

\subsection{Answering key DPIA questions}
\label{sec:DPIA}
Based on the previous paragraph, we now define six key questions relevant to the DPIA. Software development teams and DPOs should consider how to answer these questions when writing the DPIA. Each question is followed by an explanation of how our proposed analysis technique can help answer the questions.

 \begin{enumerate}[font={\bfseries},label={Q\arabic*}]
    \item \emph{What is the source \& nature of the data?}
    \Answer{We need to know where the data is acquired originally and through which way. By having privacy source methods detected from the target program, we are able to look for all the potential locations in which personal data from users might get captured by the system. Different categories of privacy source methods might also indicate the type and nature of the data. For example, a method from \texttt{java.io.File} indicates this method reads from a file in the local file system.}
    
    \item \emph{How is private data processed?}
    \Answer{We want to identify the parts of the program that involve the processing of private data. This is a discovery study based on the flows that stem from privacy source methods. There are many patterns that might provide details on the processing of privacy data, for example, data travel through multiple sources or reach into multiple different sinks.}
    
    \item \emph{Will the data be transformed? If so, how to ensure privacy data quality?}
    \Answer{Data transformation and quality control can be subtle. There are clues such as the change of data types, certain types of data manipulation methods or certain APIs that might get involved in data transformation such as encryption or database packages.}
    
    \item \emph{Will the data be shared/transferred and if yes, how?}
    \Answer{Most of the data transportation happens when the privacy data flow into a sink method. By pinpointing the location and type of sink methods, we are able to identify whether there are private data being shared or transferred out of the target program.}
    
    \item \emph{Does the data collected include special/highly sensitive personal data?}
    \Answer{The property of privacy data need to be manually identified or with the help of developers. By adopting pure logic we can pick up properties that are directly linked with specific input devices of software.}
    
    \item \emph{How is the data secured?}
    \Answer{The security of private data is ensured when there are data protection mechanisms adopted, for example, the usage of cryptographic libraries or some encrypted databases. By locating the occurrence of these methods, we are able to analyse the data security protection of the target program.}

\end{enumerate}

\section{Implementation}
\label{sec:analysis}
In the following paragraphs, we explain how our program analysis technique is implemented.
Our implementation is built on Soot~\cite{lam2011soot}, a Java optimisation framework that provides four intermediate representations for analysing and transforming Java bytecode. Our technique consists of three parts:

\begin{itemize}
    \item Transforming program bytecode to intermediate representation;
    \item Finding the source and sink methods;
    \item Building a privacy flow-graph by constructing one privacy flow for each source method at a time;
    \item Producing the abstraction extracted from the privacy flow-graph.
\end{itemize}

\subsection{Finding source and sink methods}
Soot helps us transform our target program into a 3-address intermediate representation~\cite{vallee1998jimple}. By traversing the $CFG(c.m)$ of each method $c.m$ in the program (provided in Jimple), the local data-flow analysis helps us detect the occurrences of source and sink methods in the pre-set annotation datasets ($\om$ and $\im$) defined in Section~\ref{sec:javaspec}.
By having a complete list of source and sink methods in the application as $\mathcal{O}$ and $\mathcal{I}$, we now use them to start building the privacy flow-graph.

\subsection{Building the privacy flow-graph}
For every class that includes a detected source method, we mark it as a class-of-interest (COI).
For each COI, we first build a complete call-graph for it.

\begin{definition}[Class-of-interest]
A Class-of-interest (COI) is a class that contains an invocation to one of the source methods ($\mathcal{O}$).

\begin{equation*}
    c\in \mathcal{COI} \Leftrightarrow \exists o\in c, o \in \mathcal{O}
\end{equation*}
\end{definition}

Now for each source method $o\in \mathcal{O}$, we build a global data-flow $G_o = F^o \dots F^n$ for it from the call-graphs of each class that $G_o$ passes through.
The final output is a union of all the global data-flows originating from source methods. This graph uses $A \rightarrow B$ to represent that method $B$ invokes method $A$.
Each $G_o$ will be output as a separate \verb|dot| file consisting of all the nodes (full signature of methods) and edges (invocations among the methods) which enables users to easily visualise it with simple tools.

\subsection{Abstracting the privacy flow-graph}
\label{sec:result}
Privacy flows can be lengthy and comprise a variety of non-sensitive processes, many of which are from the same class and are unrelated to privacy protection yet may confound both developers and DPOs. We want to enable DPOs to get a big picture of the important processes without getting bogged down in minutiae by creating an abstraction from the privacy-flow-graphs generated by each source method.
The abstraction is powered by a simple Python script running automatically on the initial complete privacy flow-graph.
We select several key parts from the complete privacy flow-graph which are listed below as symbols:

\begin{itemize}
    \item $\blacktriangle$: the starting source method;
    \item $\vartriangle$: a non-starting source method;
    \item $\ocircle$: a non-special process;

    Multiple processes that belong to the same package will be grouped into one process symbol in the abstraction.
    \item $\otimes$: a security process (cryptography, database, or network);

    A security process is detected by the substring detector, we look for substrings such as `encrypt', `db', `send', `connect' in the method and its package name.
    \item $\blacktriangledown$: the end sink method;
    \item $\triangledown$: a non-ending sink method;
    \item $\CIRCLE$: the end process;
    \item $\lozenge$: an authentication process;

    Similar to a security process, we report an authentication process when we detect the substring `auth' in the method or its package name.
    \item $\odot$: initialisation process(es).

    The initialisation process has `init' in their names which can be picked up by our substring detector.
\end{itemize}

The above key information can be interpreted to help developers pin down specific issues in code and assist DPOs to have a sketch of high-level privacy patterns in the program, to also better answer the relevant questions in DPIA.

An example abstraction output reflecting the code snippet in Figure~\ref{fig:javaexample} is shown below:
\begin{figure*}[ht]
    \caption{Example of a privacy data-flow generated for a source code fragment and its abstraction}
    \includegraphics[width=0.8\textwidth]{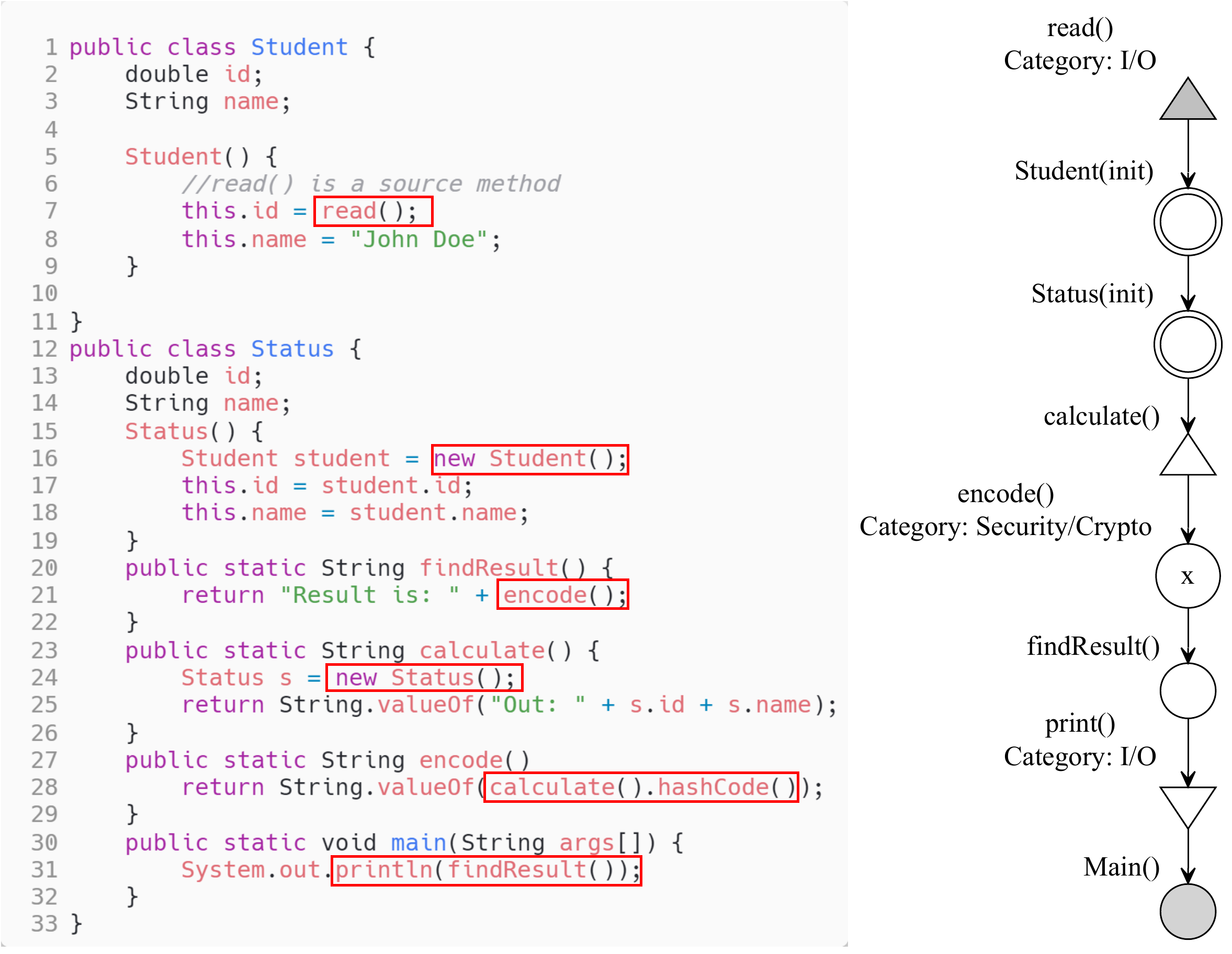}
    \label{fig:javaexample}
\end{figure*}
The example has one obvious source method \texttt{read()} (line 7) which acts as the starting point of our analysis. The technique then finds the next invocation to the source method when class \texttt{Student} gets initialised (line 16). This initialisation is triggered later by another initialisation of class \texttt{Status} (line 24). Following the newly created object \texttt{Status s}, we can trace the invocations to \texttt{calculate()} (line 28), \texttt{encode()} (line 21), \texttt{findResult()} (line 31) and finally to a sink \texttt{print()} (line 31) which is invoked by the \texttt{Main()} method. Source method \texttt{read()} and sink method \texttt{print()} have their categories labelled as well as the special process \texttt{encode()}.

Along with the abstraction figure, we provide short labels with the symbols which consist of information such as 1) categories of starting source method and sink methods; 2) categories of the special processes (security, authentication, or initialisation); 3) the class name is displayed when it is an initialisation process (optional).

\section{Experiment}
\label{sec:experiment}

We are looking for apps that accept raw sensitive user data and entail data transmission, often in messaging and cloud storage applications. We thus selected the following two applications: Signal\footnote{\url{https://signal.org/en/}} and NextCloud\footnote{\url{https://nextcloud.com/}}. The non-profit Signal Foundation and Signal Messenger LLC created Signal, a cross-platform end-to-end instant messaging service. We intend to study how Signal processes privacy-related user data by analysing both Signal's front-end Android application and the Signal Client Service API because of its expertise in end-to-end encryption. The purpose is to figure out how data is taken from the user and sent to the server. NextCloud is a client-server software package for developing and managing file hosting services. It is free and open-source software that anybody may install and run on their own private servers. We chose an implementation of its Client API that assists developers in developing Java apps with NextCloud integration since it is highly configurable. Similar to Signal, we intend to determine how the application handles privacy-sensitive user data.

\subsection{Signal}
The Signal Service API contains 17,710 lines of code, which might require developers and DPOs significant time and effort to comprehend. With our samples of DPIA answers, DPOs could effortlessly use our implementation results to create a DPIA.

\begin{figure*}
\caption{Sample abstract privacy flows for Signal and NextCloud\label{fig:result}}
\includegraphics[width=.9\textwidth]{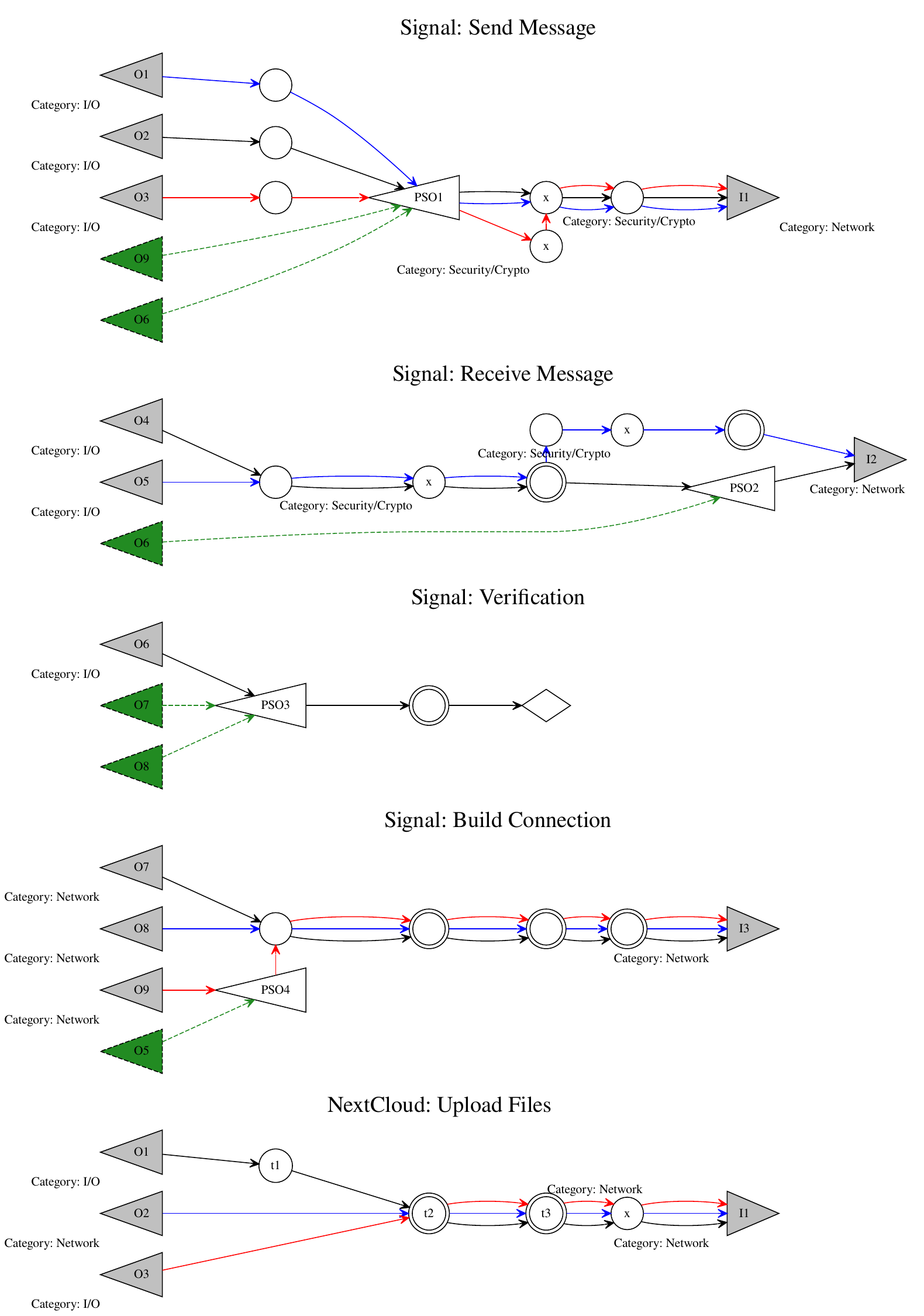}
\end{figure*}

A total number of 11 privacy flows were detected in Signal Service API (9 out of a total 11 are displayed here), the abstraction of its privacy flow-graph is shown below as Figure~\ref{fig:result}. We categorise the 9 source methods found into four 4 different functionalities. 
In Signal, we have discovered a similar pattern for all types of data communication: each raw entry is instantly sent into Signal's own cryptography libraries, allowing all user entries to be completely encrypted before they reach any possible sinks or processes. \textit{Signal: Send Message} and \textit{Signal: Receive Message} in Figure~\ref{fig:result} demonstrate this end-to-end encryption mechanism.
As indicated by the dashed green lines, there are some source methods that accept some values from local fields which originated from source methods in other flows. $\texttt{PSO1}$, for example, gets value from source methods $\texttt{O6}$ and $\texttt{O9}$, which are network-related properties associated with the message object.

Now, we answer the DPIA questions we listed in Section~\ref{sec:DPIA} using the flow that originates from \texttt{O1} (blue flow) in \textit{Signal: Send Message} of Figure~\ref{fig:result}.
To analyse privacy compliance, we combine the abstraction figure (which only comprises shapes and categories of critical processes) with detailed privacy flow-graphs (which contain every node in the flow-graph as well as their complete signature), shown as in Table~\ref{tab:detailed}.

\begin{table*}[ht]
\centering
\caption{Complete privacy data-flow with abstraction symbols for sending a text message in Signal}
\label{tab:detailed}
\resizebox{\textwidth}{!}{%
\begin{tabular}{|c|l|}
\hline
\textbf{\textbf{Abstraction}} & \multicolumn{1}{c|}{\textbf{Complete privacy data-flow}}                                                                 \\ \hline
$\blacktriangle$              & android.widget.EditText getText()                                                                                        \\ \hline
$\ocircle$                    & org.thoughtcrime.securesms.jobs.PushTextSendJob deliver(message)                                                         \\ \hline
$\vartriangle$                & org.thoughtcrime.securesms.messages.MessageContentProcessor handleMessage(content, timestamp, ...)                       \\ \hline
$\otimes$                     & org.whispersystems.signalservice.api.crypto.SignalServiceCipher encrypt(destination, message, ...)                       \\ \hline
\multirow{3}{*}{$\ocircle$} & org.whispersystems.signalservice.api.SignalServiceMessageSender getEncryptedMessage(content, recipient, timestamp, ...) \\ \cline{2-2} 
                              & org.whispersystems.signalservice.api.SignalServiceMessageSender getEncryptedMessages(content, recipient, timestamp, ...) \\ \cline{2-2} 
                              & org.whispersystems.signalservice.api.SignalServiceMessageSender createMessageContent(message)                            \\ \hline
$\blacktriangledown$          & org.whispersystems.signalservice.api.SignalServiceMessageSender sendMessage(message, recipient, ...)                     \\ \hline
\end{tabular}%
}
\end{table*}

\begin{enumerate}[font={\bfseries},label={Q\arabic*}]

\item \emph{What is the source \& nature of the data?}
\Answer{Android applications take text input from a TE object which is a UI fragment providing a text field for users. The message field contains the raw message users want to send out.}

\item \emph{How is private data processed?}
\Answer{The abstraction tells us that there exist multiple processes when the text message is being sent out. There are two non-privacy processes from packages \texttt{org.signal.securesms.jobs} and \texttt{org.signalservice.api.signalservicemessagesender}. The package names indicate the types of processing behind the processes. There are also highly sensitive privacy processes such as the \texttt{MessageContentProcessor()} which is a non-starting source method that takes privacy data from a local field, in this case, it combines multiple privacy data including the text message. \texttt{org.signalservice.api.crypto} shows a typical encryption process, this also demonstrates the end-to-end encryption in Signal.}

\item \emph{Will the data be transformed? If so, how to ensure privacy data quality?}
\Answer{We notice that the data type gets immediately changed after being read into the device as raw strings. Both non-privacy and privacy processes transform data in order to achieve their functionality. However, encrypted messages stay encrypted before they get sent out, which ensures the content will not get manipulated by external parties.}

\item \emph{Will the data be shared/transferred and if yes, how?}
\Answer{The final ending sink method \texttt{sendMessage()} sends encrypted message objects out to the server from the client.}

\item \emph{Does the data collected include special/highly sensitive personal data?}
\Answer{The properties of the message object are sensitive. Not only the text message body itself, its attributes such as the details of senders but receivers and timestamps also remain sensitive during the entire process.}

\item \emph{How is the data secured?}
\Answer{Data security is guaranteed here by end-to-end encryption. All the privacy data related to the message get encrypted together as an \texttt{EncryptedMessage} object. This encrypted object cannot be decrypted by the server, which remains unreadable until it reaches the destination client.}
\end{enumerate}

Our discovery also supports what Signal claims in its privacy policy. By supplying the aforesaid information to both developers and DPOs, they are able to receive adequate information for creating DPIA and examining the privacy protection status in Signal without having to read the original code.

\subsection{NextCloud}
Since NextCloud recently implemented end-to-end encryption in their products, this feature only offers on the level of `end-to-end encrypted folders'. Hence in our analysis, we only apply the technique to the client API which is applied to the traditional version that relies on TLS communication for safely transferring files.

From a total of 8,923 lines of code, we are able to extract key information from the NextCloud Client API using a simplified privacy flow-graph along with the complete flow-graphs with full signatures, as we did with Signal. We evaluate the DPIA questions to help DPOs in getting information from a legal standpoint, using the abstraction graph derived from our technique in Figure~\ref{fig:result}.

\begin{enumerate}[font={\bfseries},label={Q\arabic*}]

\item \emph{What is the source \& nature of the data?}
\Answer{NextCloud Client API allows a client to upload a new file via \texttt{uploadNewFile()}. The files can be of various types but shall be categorised as the user's personal data. There is also one network source, which links with data that can be used to identify users on the Internet.}

\item \emph{How is private data processed?}
\Answer{The file is transmitted from the device to the network; this is how a file is sent from the client to the server.}

\item \emph{Will the data be transformed? If so, how to ensure privacy data quality?}
\Answer{Not only the file acquired from the user is transferred to the server, but also network data and configuration settings. These various user data are processed and loaded into multiple fields of various class objects (reflect on the two initialisation processes). During these procedures, data types must be transformed in order to be organised for transmission as a type that the server accepts.}

\item \emph{Will the data be shared/transferred and if yes, how?}
\Answer{The final node is a network sink, which indicates that the user's data has been transmitted into the network and shared with the server.}

\item \emph{Does the data collected include special/highly sensitive personal data?}
\Answer{In this example, the data comprises user files, settings, and network details. User files are highly sensitive in terms of privacy.}

\item \emph{How is the data secured?}
\Answer{The network process here depicts a TLS connection, which is a cryptographic technology meant to ensure network communications security.}
\end{enumerate}
With the information provided above, we provide both developers and DPOs a better understanding of how the file upload process works in the NextCloud Client API, as well as what and where are the important aspects of privacy protection for NextCloud.

Privacy flow-graphs illustrate trends in terms of privacy-related data processing, including both benign and bad practices. It can assist not just DPOs and developers in responding to DPIA questions and addressing important processing, but also in identifying potentially questionable practices and ensuring good practices on privacy-related data.

\section{Related Work}
Using static analysis for security bug detection in software~\cite{chess2004static,ayewah2008using,evans2002improving} is a source of inspiration for our work. In our work, we used hand-crafted datasets of source and sink methods for Java and popular third-party libraries as the start point for our analysis. The idea of using a pre-built set as a basis of static analysis is similar to SUSI~\cite{arzt2013susi}, IccTA~\cite{li2015iccta}, MudFlow~\cite{7194594} and AndroidLeak~\cite{gibler2012androidleaks} in terms of privacy protection for Android applications. Most current work, including the above, is specific to Android sinks and sources and often uses name features as the basis of their analysis, whereas we focus on Java in general without adopting heuristics. Regarding the GDPR, we demonstrate the utility of employing privacy flow-graphs to ease the DPIA process, which saves manual labour and assists in identifying possible sensitive processes that may be missed by human eyes.

Overall, there is an increasing interest in assuring privacy protection compliance prior to or throughout the software development lifecycle~\cite{rubinstein2011regulating}. Privacy-by-design (PbD) has sparked research into methodologies and models for preserving software privacy before implementation begins, as well as forecasting or managing developer privacy compliance throughout implementation~\cite{hadar2018privacy,antignac2014privacy,hoepman2014privacy}.
Many of these approaches may also be employed on a regular basis during the development cycle and while updating software. In the era of GDPR in Europe, there is also prior research~\cite{horak2019gdpr,bu2020controller,henriksen2020dpia} that aims to provide personalised solutions for DPIA in a variety of applications. According to a survey conducted by Dias Canedo \emph{et al.}~\cite{dias2020perceptions}, technical staff frequently lack legal knowledge regarding privacy protection. Many existing works~\cite{8406568,massey2010evaluating,piras2019defend} propose models that limit on a conceptual level, that are not tangible for both technical and non-technical people to apply to implementation, motivating us to propose an automatic technique to analyse privacy compliance in software.

\section{Conclusion}
In terms of privacy protection, there always exists a barrier between developers and DPOs. DPOs need to generate a successful DPIA to document the privacy protection behaviour of software, this requires the developer's comprehensive knowledge of code details.
Our work provides a technique for detecting privacy source and sink methods in software bytecode, generating privacy flow-graphs from the discovered sources, and supporting DPOs in writing a DPIA utilising privacy flow-graphs and associated abstractions.

\section{Limitation and future work}
Our present method requires predetermined source and sink lists.
Given that modern applications typically contain hundreds of direct and indirect dependencies, we may miss a significant number of privacy-related sources and sinks.
Therefore, we rely on the knowledge of technical specialists to create a more precise list of sources and sinks.
Moreover, despite the fact that our complete privacy flow-graphs and their abstractions can express key privacy-sensitive behaviours such as data acquisition, encryption, and transportation, they are unable to provide complete information regarding which type of data manipulation was involved in terms of privacy protection; therefore, developers may be required to provide additional explanation for DPOs.

Future work includes a more detailed local flow analysis for each local data-flow in a privacy global data-flow, such as tracking how values from privacy-related data are modified in the local method and flagging sensitive manipulations such as value accumulation and separation. 
In the meantime, it is feasible to extract information from the manifest file on which third-party libraries are imported by the software in order to assist in the construction of a more adaptable list of sources and sinks.
This procedure might be automated by including these third-party libraries (which are usually downloadable as JAR files) as a part of the input of the analysis.
Additionally, since dynamically-typed languages such as JavaScript are used in many different types of modern systems, it would be advantageous to build a source code-based analyser based on tools such as Semgrep \footnote{\url{https://semgrep.dev/}}, which as a starting point for extending our results to web applications.
\begin{acks}
We appreciate the legal insight that Jan Czarnocki and Lydia Belkadi have given. This work is part of the Privacy Matters (PriMa) project. The PriMa project has received funding from European Union’s Horizon 2020 research and innovation program under the Marie Skłodowska-Curie grant agreement No. 860315.
\end{acks}


\bibliographystyle{ACM-Reference-Format}
\bibliography{references}

\end{document}
\endinput